\documentclass[aps,prx,preprint,unsortedaddress, longbibliography]{revtex4-1}
\usepackage{graphicx}
\usepackage{color}
\usepackage{amsmath}
\usepackage{lineno}
\usepackage{float}
   \usepackage{setspace}

\begin{document}
\title{Gigahertz free-space electro-optic modulators based on Mie resonances}



\author{Ileana-Cristina Benea-Chelmus}%
\email{cristinabenea@g.harvard.edu}
\affiliation{Harvard John A. Paulson School of Engineering and Applied Sciences, Harvard University, Cambridge, MA, USA}
\author{Sydney Mason}
\affiliation{Harvard College, Cambridge, MA, USA}
\author{Maryna L. Meretska}
\affiliation{Harvard John A. Paulson School of Engineering and Applied Sciences, Harvard University, Cambridge, MA, USA}
\author{Delwin L. Elder}
\affiliation{University of Washington, Department of Chemistry, Seattle, WA, USA}
\author{Dmitry Kazakov}
\affiliation{Harvard John A. Paulson School of Engineering and Applied Sciences, Harvard University, Cambridge, MA, USA}
\author{Amirhassan Shams-Ansari}
\affiliation{Harvard John A. Paulson School of Engineering and Applied Sciences, Harvard University, Cambridge, MA, USA}
\author{Larry R. Dalton}
\affiliation{University of Washington, Department of Chemistry, Seattle, WA, USA}
\author{Federico Capasso}
\email{capasso@seas.harvard.edu}
\affiliation{Harvard John A. Paulson School of Engineering and Applied Sciences, Harvard University, Cambridge, MA, USA}

\date{\today}

\newpage

\begin{abstract}
Electro-optic modulators from non-linear $\chi^{(2)}$ materials are essential for sensing, metrology and telecommunications because they link the optical domain with the microwave domain. At present, most geometries are suited for fiber applications. In contrast, architectures that modulate directly free-space light at gigahertz~(GHz) speeds have remained very challenging, despite their dire need for active free-space optics, in diffractive computing or for optoelectronic feedback to free-space emitters. They are typically bulky or suffer from much reduced interaction lengths. Here, we employ an ultrathin array of sub-wavelength Mie resonators that support quasi bound states in the continuum~(BIC) as a key mechanism to demonstrate electro-optic modulation of free-space light with high efficiency at GHz speeds. Our geometry relies on hybrid silicon-organic nanostructures that feature low loss~($Q = $~550 at $\lambda_{res} = 1594$~nm) while being integrated with GHz-compatible coplanar waveguides. We maximize the electro-optic effect by using high-performance electro-optic molecules (whose electro-optic tensor we engineer in-device to exploit $r_{33} = 100$~pm/V) and by nanoscale optimization of the optical modes. We demonstrate both DC tuning and high speed modulation up to 5~GHz~($f_{EO,-3~dB}  =3$~GHz) and shift the resonant frequency of the quasi-BIC by $\Delta\lambda_{res}=$11~nm, surpassing its linewidth. We contrast the properties of quasi-BIC modulators by studying also guided mode resonances that we tune by $\Delta\lambda_{res}=$20~nm. Our approach showcases the potential for ultrathin GHz-speed free-space electro-optic modulators. 
\end{abstract}
\maketitle

\section{Introduction}
\begin{figure}[htb]
  \centering 
 \includegraphics[width=16cm]{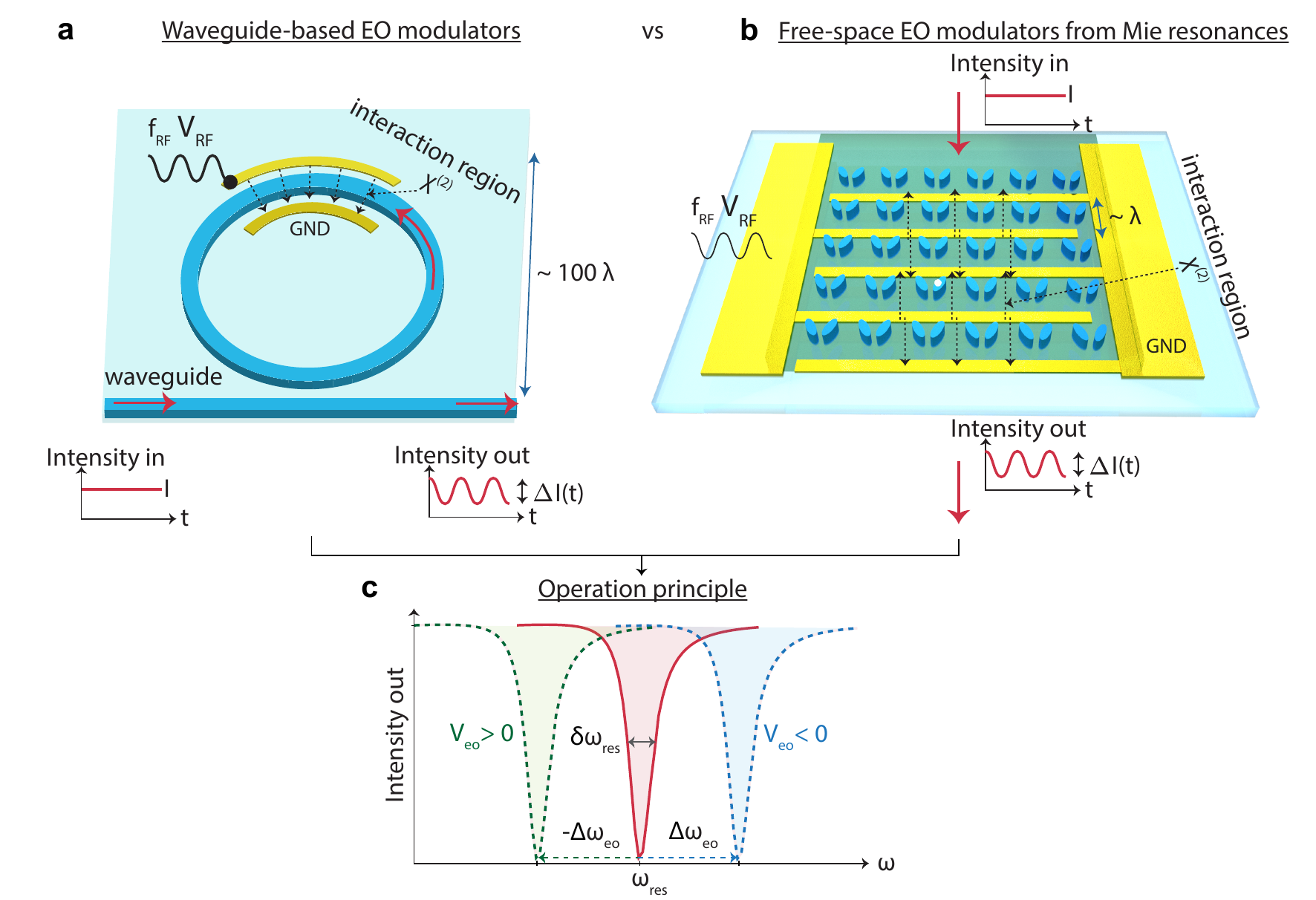}
  \caption{\textcolor{black}{\textbf{Comparison of waveguide-based and free-space electro-optic modulators.} \textbf{a,} Waveguide-based electro-optic modulators rely on resonant structures that are in on-chip waveguides such as e.g. ring resonators of high azimuthal order or on-chip interferometers. Light propagating in the waveguide acquires a phase modulation at the RF frequency $f_{RF}$ due to the electro-optic effect; the interaction region may be hundreds of wavelengths long. \textbf{b,} In contrast, free-space electro-optic modulators change the properties of a beam that is incident from free space. Sub-wavelength Mie nanoresonators impart a phase modulation via the electro-optic effect to the incident light that propagates through the thin film; the interaction length is typically shorter than a single wavelength. \textbf{c,} Resonant electro-optic modulators work on the principle that their resonant frequency $\omega_{res}$ is tuned by $\Delta \omega_{eo} (t)$ linearly by an applied bias, due to the phase shift induced by the electro-optic effect. A radio-frequency bias $V_{RF} (t)= V_{eo}\times sin(2\pi f_{RF} t) $ displaces the resonance frequency around its zero bias value. Narrowband resonances that satisfy $\Delta \omega_{eo} >  \delta \omega_{res}$ are preferred for full intensity modulation at low switching voltages. Dashed black arrows indicate the applied tuning field that introduces the electro-optic effect. Red arrows indicate the propagating optical field. EO = electro-optic, GND = ground.}}
  \label{fig:FigVision}
\end{figure}

Recently, photonic technologies have become promising to address the bottleneck for high-speed communication~\cite{ShiTianGervais+2020+4629+4663} and high-performance computing~\cite{Psaltis1990,Zhou2021} instead of traditional all-electronic technologies. The next-generation photonic devices need to manipulate light in multiple independent channels at high speeds, and metasurfaces are inherently suited to provide spatial multiplexing capabilities~\cite{Zheludev2012,Shaltouteaat3100}. Still today, the vast majority of photonic systems are passive. Amongst the available mechanisms that are explored for active control of light, hybrid~\cite{Monat2020} electro-optic interconnects employ $\chi^{(2)}$ effects to transduce a signal from the electronic domain into the optical domain and combine the low-loss, low-dissipation hallmarks of photonics with the compactness and reconfigurability of electronic circuits. Importantly, they have proven superior to alternative techniques when it comes to speed: the control fields may oscillate at frequencies well beyond the microwaves into the terahertz~\cite{Benea-Chelmus:20}. As a result, an unprecedented variety of electro-optic modulators has been reported, driven by the progress in the molecular engineering~\cite{D0TC05700B}, growth~\cite{Eltes2016}, fabrication and stability~\cite{Lu2020} of materials such as organic non-linear molecules~\cite{Xu2020}, barium titanate~\cite{Abel2019a} and lithium niobate~\cite{Luke:20}. Today, most demonstrations target fiber applications and foster photonic integrated platforms~\cite{Zhu:21, Elshaari2020} such as silicon-on-insulator technologies, where a low optical loss and a high transduction efficiency have been achieved. These two figures of merit are key for a wide range of applications, and are especially crucial for quantum applications~\cite{Rueda2019a, Youssefi2021}.  Waveguide-based electro-optic modulators have a size on the order of few tens to few hundreds of wavelengths~(such as e.g. ring resonators of the kind shown in Fig.~\ref{fig:FigVision}~a), since it allows one to use whispering gallery modes with lowest radiative losses and hence narrowest linewidth.

Instead, ultrathin electro-optic modulators from sub-wavelength resonators are exquisite candidates in applications that require direct control over free-space light in a compact and spatially multiplexed way, such as free-space optical communication links~\cite{Mphuthi:19}, coherent laser ranging, active optical components, high-speed spatial light modulators~\cite{Smolyaninov2019,Benea-ChelmusI.-C.MeretskaM.ElderL.D.TamagnoneM.DaltonR.L.Capasso2021} and active control of free-space emitters~\cite{Traverso:21}. Flat optical components such as metasurfaces~\cite{Chen2020,Khorasaninejad1190} rely on sub-wavelength sized nanostructures that change the properties of a beam that is incident from free-space onto the metasurface. In contrast to waveguide-based modulators, free-space electro-optic modulators from $\chi^{(2)}$ materials have been researched less~(few examples are Refs.~\cite{Gao2021,Timpu2020,doi:10.1002/adom.202000623}) despite their commonalities illustrated in Fig.~\ref{fig:FigVision}~a and b. In any electro-optic modulator, the the microwave field is applied via metallic electrodes~\cite{Wang2018, Haffner2018} or antenna structures~\cite{Salamin2015, Benea-Chelmus:20} and changes the refractive index $n_{mat}$ of the non-linear material at optical frequencies via the linear electro-optic effect, also known as Pockels effect, by $\Delta n (t) = - \frac{1}{2}n_{mat}^3rE (t)$, with $r$ the electro-optic coefficient of the material and $E (t)= \frac{V_{RF}(t)}{d}$ the tuning field~(voltage $V_{RF}(t)$ applied across the distance $d$). In resonant modulators that employ the $r_{33}$ component of the electro-optic tensor, this change in refractive index $\Delta n(t)$ modifies the resonant frequency as illustrated in Fig.~\ref{fig:FigVision}~c by $\Delta \omega_{eo}(t) =  -\frac{\Delta n (t)}{n_{mat}}\omega_{res}\Gamma_c = g_{eo}V_{RF}(t)$, with $g_{eo} = \frac{1}{2}n_{mat}^2r_{33} \frac{1~V}{d}\omega_{res}\Gamma_c$ the electro-optic coupling rate at 1~V applied voltage~\cite{Zhang2019} and $\Gamma_c$ the overlap factor of the two interacting fields with the nonlinear medium. The shift is proportional to the applied voltage and its polarity. 

The fundamental challenge to realize free-space electro-optic modulators as shown in Fig.~\ref{fig:FigVision}~b compared to waveguide-based electro-optic modulators stems, preponderantly, from the quality factors that can be achieved in the two systems and thus the underlying available interaction times.  Resonances with a high quality factor~(and thus a small full width half maximum $\delta \omega_{res} = 2\pi\times \delta f_{res}$) are favorable as they minimize the so-called switching voltage $V_{eo} = V_{switch}$ that is necessary to fully shift the resonance away from its unbiased value, which occurs when $\delta \omega_{res} \leq \Delta \omega_{eo}$. In conditions of high-Q, an optical beam experiences full modulation of its intensity or phase even for low $V_{eo}$. The frequency shift can be derived from the phase modulation $\Delta \phi_{eo} = \Delta \omega_{eo}t_{int}$ introduced by the Pockels effect, where $t_{int} = \frac{2\pi}{\gamma_{rad}}=\frac{2\pi}{\delta \omega_{res}} $ is the interaction time of the optical beam with the control field within the nonlinear material and $\gamma_{rad}$ the radiative loss rate of the optical field out of the interaction region, into the far-field.  The natural wavelength-scale dimensions of free-space modulators typically limits the efficiency straight in two ways. First, the spatial extent of the interaction region is only few hundreds of nanometers long, commensurate with the typical thickness of flat optics. This results into an electro-optic transduction that is often inefficient. Second, wavelength-sized resonators have long been contended with low quality factors as a result of their small azimuthal modal order.  However, recent breakthroughs in engineering of high-Q plasmonic resonators~\cite{Bin-Alam2021} or Mie resonators~\cite{Koshelev2021} from silicon nanoantennas~\cite{Lawrence2020} or bound states and quasi-bound states in the continuum~(quasi-BICs)~\cite{PhysRevLett.121.193903} have showcased compelling free-space candidates that now routinely reach quality factors on the order of few hundreds to few thousands, albeit in electronically passive structures that do not necessitate integration with microwave metallic waveguides that can introduce significant losses and lower the performance. 

In this work, we harness quasi-BIC resonances for hybrid silicon-organic electro-optic modulators that feature a small footprint and low dimensions (as illustrated in Fig.~\ref{fig:FigVision}~b) and that preserve a Q-factor up to 550 even when homogeneously integrated with high-performance electro-optic molecules and interdigitated driving electrodes.  A highly efficient electro-optic transduction is made possible by state-of-the-art $\chi^{(2)}$ organic molecules JRD1 in polymethylmethacrylare~(PMMA)~\cite{doi:10.1063/1.4884829} that are low-loss and are spatially located within the high-field areas of the optical nearfield. By judicious three-dimensional engineering, we incorporate metallic coplanar waveguides that provide GHz-speed driving fields.  In short, the physics behind quasi-BICs relies on confined modes that can exist inside a continuum. Our geometry~(discussed in detail in~\cite{PhysRevLett.121.193903}) explores symmetry breaking as a mean to influence the linewidth of the optical modes. Finally, we benchmark the performance of quasi-BIC modes for free-space transduction against guided mode resonances~(GMR) that can arise in similar nanostructures. GMRs appear due to the scattering of incident light by the silicon pillars into grating orders that correspond to the propagation vector of guided modes.  

\section{Results}

An array of elliptical silicon resonators is patterned on a quartz substrate and gold interdigitated electrodes are deposited around the resonators and then covered by the high-quality active organic layer. The embedded array has a sub-wavelength thickness and operates in a transmission geometry where an optical field is normally incident from free-space. Fabricated devices are shown in Fig.~\ref{fig:FigDevice} and the fabrication protocol is discussed in the methods and sketched in Supplementary Fig.~S1. The scanning electron micrographs~(SEM) of Fig.~\ref{fig:FigDevice}~a, c and d  show the array of silicon resonators prior to and after the deposition of the metallic electrodes. The layer of organic electro-optic molecules consists of JRD1:PMMA of $50\%$wt, has a lower refractive index than silicon ($\tilde n = n+ik$, with $n = 1.67$ and $k = 5\times 10^{-5}$ at $\lambda_{res}=1550~nm$) and is not shown here. Wavelength-, voltage- and concentration-dependent properties of the active layer are extensively reported in reference~\cite{Benea-ChelmusI.-C.MeretskaM.ElderL.D.TamagnoneM.DaltonR.L.Capasso2021} and its associated supplementary information. By choice of the exact geometrical parameters of one unit cell of the array illustrated in the insets of Fig.~\ref{fig:FigDevice}~a~(TV = top view) and d~(SV = side view), the silicon nanostructures can be engineered to exhibit quasi-BIC and guided mode resonances in the C- and L- telecom bands, as shown in Fig.~\ref{fig:FigDevice}~b. While their linewidth is clearly very distinct, both types of optical modes are localized mostly in the layer of organic molecules and outside the high-index silicon material, as demonstrated by their field profiles in Fig.~\ref{fig:FigDevice}~e and f, where the arrows denote the orientation of the electric field in the plane A of the resonators.  Moreover, we chose to pattern the silicon resonators of height $h_{Si} = 200~$nm on top of an elliptical pedestal from silicon dioxide of height $h_{SiO_2}=$200~or~300~nm, as visible from both the SEM figures and the cross-section of one unit cell shown in Fig.~\ref{fig:FigDevice}~b. This step is essential to minimize the overlap and thereby the losses of the optical field with the metallic electrodes. Simulated cross-sections of the optical field are displayed in Supplementary Fig.~S2 to demonstrate its localization in the near-field of the silicon resonators. 

We note a particular feature of the two modes considered here. While both resonances are excited with x-polarized light, the optical nearfield of the resonators is mainly z-polarized for the quasi-BIC mode and x-polarized for the GMR mode. This fact explains our choice of electrode orientation that is different for the two resonances: for the quasi-BIC structure, the electrodes are parallel to the x-axis, while for the GMR structure, they are parallel to the z-axis. This orientation maximizes the alignment of the optical nearfield parallel to the applied RF-field, oriented perpendicularly to the electrodes, and allows us to exploit the $r_{33}$ coefficient of the electro-optic tensor of the JRD1:PMMA layer. We note that in the case of the organic layer used here, the orientation of the electro-optic tensor with respect to the geometrical coordinate system of the sample is established post-fabrication, by electric field poling, a procedure during which the molecules orient along a DC electric field~\cite{Benea-ChelmusI.-C.MeretskaM.ElderL.D.TamagnoneM.DaltonR.L.Capasso2021, Heni2017a,Xu2020} that is applied via the gold electrodes. By definition, in the organic layer utilized here, $r_{33}$ corresponds to the direction of the poling field. We provide in Supplementary Fig.~S2 electrostatic simulations of the poling fields for the two geometries. The orientation of the poling fields and their relative strength is indicative of the orientation and level of alignment of the organic electro-optic molecules. Since in our case, the entire array is poled at once by interdigitated electrodes, the electro-optic coefficient $r_{33}$ alternates in sign from one electrode period to the next as illustrated by the green and red areas in Fig.~\ref{fig:FigDevice}~c~-~d yielding an overall in-plane periodically poled JRD1:PMMA film with a typical $r_{33} = 100$~pm/V as was previously demonstrated and characterized in detail in reference~\cite{Benea-ChelmusI.-C.MeretskaM.ElderL.D.TamagnoneM.DaltonR.L.Capasso2021}. As a result, this particular structure allows to maximize the overlap factor $\Gamma_c$ for both modes. 

\begin{figure}[htp]
  \centering 
 \includegraphics[width=16cm]{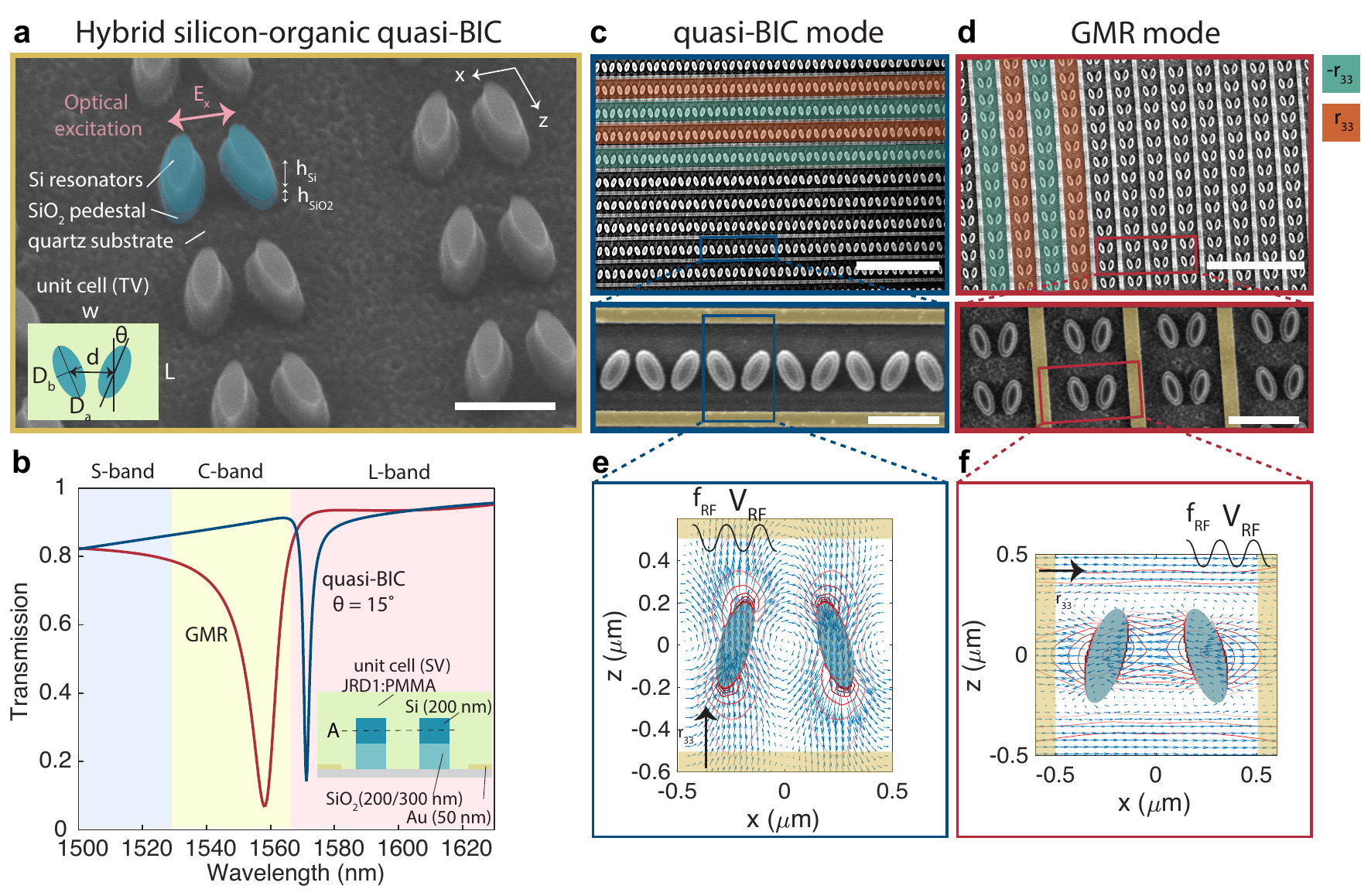}
  \caption{\textcolor{black}{\textbf{Hybrid silicon-organic free-space electro-optic modulators based on Mie resonances for C- and L-bands.} \textbf{a,} A single electro-optic modulator is made from a rectangular array of silicon nanoresonators patterned onto a quartz substrate on top of a silicon dioxide pedestal, here shown prior to the deposition of the metallic electrodes and the organic electro-optic layer~(green) which is applied post-fabrication by spin-coating and covers the nanoresonators. Inset shows the top view~(TV) of one single unit cell. Scalebar = 500~nm. This geometry can sustain two distinct types of resonances, quasi-bound states in the continuum~(quasi-BICs) and guided mode resonances~(GMR), shown in \textbf{b,}, with corresponding geometries as in \textbf{c,} and \textbf{d,}. Inset shows the side-view~(SV) of one unit cell. The two types of resonances are excited by an incident beam that is x-polarized and have distinct distributions of the near-fields of the resonators, shown in \textbf{e,} and \textbf{f,} (cross-section A of SV). While the quasi-BIC mode is circulating in the nearfield, and has a dominant component along the z-axis~(hence perpendicular to the excitation polarization), the guided mode resonance is predominantly x-polarized~(as the excitation). Given this vectorial orientation of the optical fields in the near-field of the resonators, metallic electrodes are deposited in between each row of ellipses and are oriented along the x-axis for the quasi-BICs and along the z-axis for the GMR, as shown in \textbf{b,} and \textbf{c,}~(scale bars upper picture = $5~\mu$m and lower pictures = $1~\mu$m). The interdigitated electrodes serve for the activation of the JRD1:PMMA layer by electric field poling and for the application of DC and RF tuning fields. Black arrows indicate poling direction.}}
  \label{fig:FigDevice}
\end{figure}

\begin{figure}[htp]
  \centering 
 \includegraphics[width=14cm]{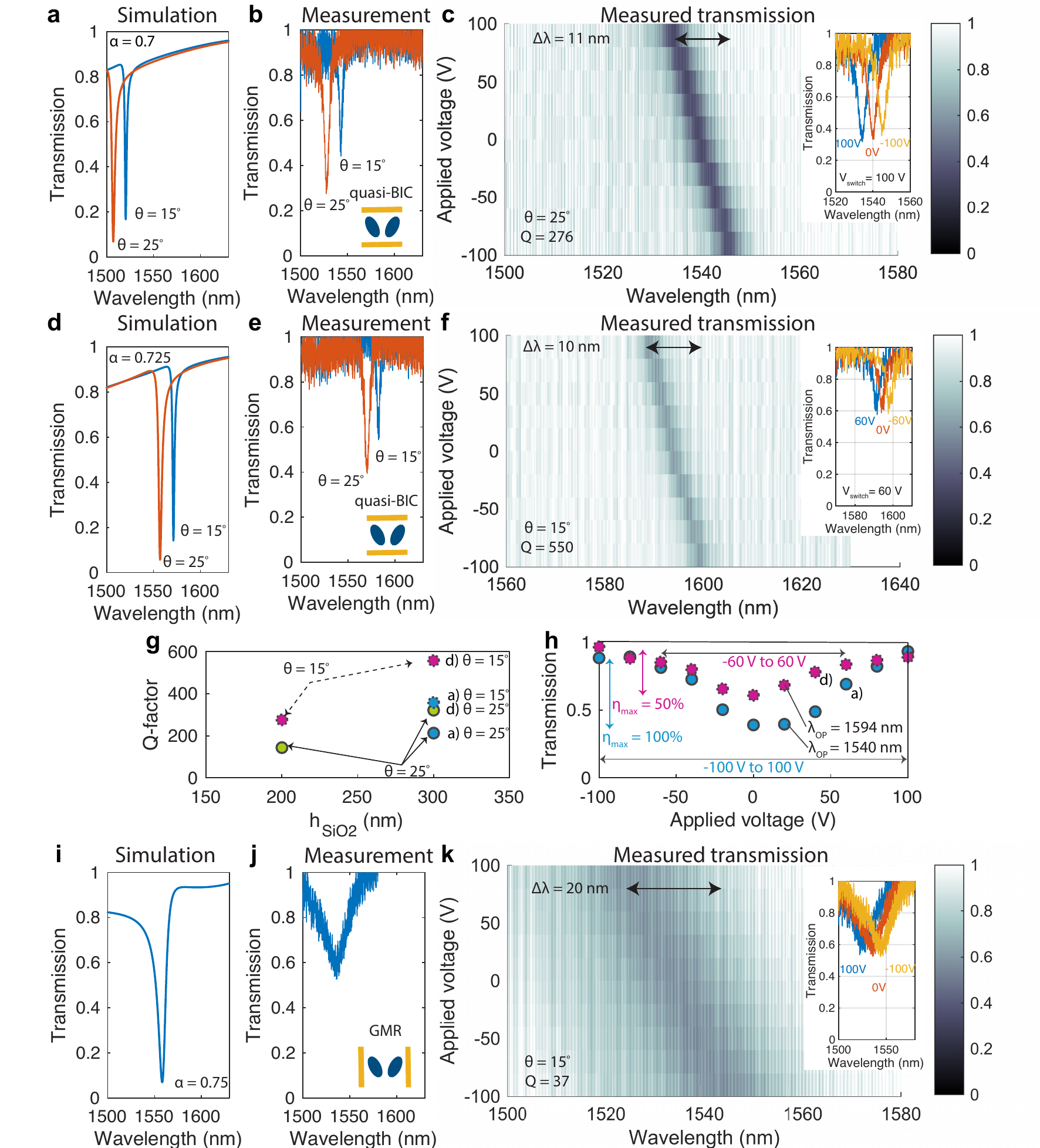}
  \caption{\textcolor{black}{\textbf{DC tuning properties of Mie-based modulators.} \textbf{a,-b,} and \textbf{d,-e,} Experimental transmission results are compared to simulated transmission curves for various geometries of electro-optic modulators based on quasi-BIC structures. We find, as expected, by experiment and simulation, that the geometrical scaling factor $\alpha$ shifts the resonances within the telecom band. In addition, the assymetry angle $\theta$ influences the linewidth of the resonances. \textbf{c,} and \textbf{f,} are DC tuning maps of the electro-optic modulators for $(\alpha , \theta) = (0.7, 25^\circ)$ and  $(\alpha , \theta) = (0.725, 15^\circ)$, respectively. Insets show three distinct curves at 0~V, and $V_{switch} = \pm100~$V and $\pm60~$V, respectively. \textbf{g,} Experimentally extracted quality factors for two distinct heights of the silicon dioxide pedestal are compared and we find that an increase in height from 200~nm to 300~nm leads to an increase of the quality factor. Dashed circles represent quasi-BIC structures with $\theta = 15^\circ$ and full contour circles represent $\theta = 25^\circ$. The circles for $h_{SiO2} = 300~$nm denote the measurements as-labled to the right of the circles. The circles for $h_{SiO2} = 200~$nm represent measurements of equivalent structures with $h_{SiO2} = 200~$nm. \textbf{h,} Detailed voltage-dependent transmission curves are reported for two exemplary operating wavelengths for the case of the two device geometries discussed in \textbf{a,} and \textbf{d,}. Full switching between ON-OFF-ON states of transmission is achieved for both geometries. \textbf{i,}-\textbf{j,} In contrast, GMRs in the same structure have much broader linewidths, demonstrated by experiment and simulation. \textbf{k,} Their resonance wavelength can be tuned over $\Delta \lambda_{res} = 20~$nm. Q-factors and asymmetry angle $\theta$ are indicated for all colormaps.}}
     \label{fig:FigResults}
\end{figure}

In the following, we first present experimental tuning properties of the hybrid silicon-organic free-space modulators when a DC voltage $V_{eo}$ is applied to the interdigitated electrodes uniformly across the entire array ($f_{RF} = 0$). In Fig~\ref{fig:FigResults}~a~-~f we report the experimental results for operation on a quasi-BIC mode engineered in the C- or L- telecom band with geometrical dimensions characterized by a scaling parameter $\alpha$, as provided in Methods. A particular feature of the quasi-BIC mode is that its Q-factor is highly dependent on the asymmetry angle $\theta$: in the absence of material losses, the quality factor increases towards infinity in the limit of $\theta = 0$. We report in Supplementary Fig.~S3 the simulated dependence of the resonance on $\theta$ for the fabricated structures. Noting that in the presence of losses, at high Q-factors the resonance depth also decreases~(eventually leading to less intensity modulation), we choose $\theta = 15^{\circ}$ and $25^{\circ}$. We confirm by experiment the expected increase of quality factor when reducing the angle from $\theta = 25^{\circ}$~($Q = 212$ and $Q = 320$ for $\alpha = 0.7$ and $\alpha = 0.725$, respectively) to $\theta = 15^{\circ}$~($Q = 357$ and  $Q = 557$ for $\alpha = 0.7$ and $\alpha = 0.725$, respectively), see Fig.~\ref{fig:FigResults}~g. The measurements were performed on structures similar to Fig.~\ref{fig:FigDevice}~c prior to electric field poling of the devices. Furthermore, the measured red-shift of the resonance with decreasing $\theta$ are well reproduced by our simulations. In Fig.~\ref{fig:FigResults}~c we report the DC tuning characteristics of the modulator based on quasi-BIC modes as a function of applied voltage $V_{eo}$ of the structure with $\alpha = 0.7$ and $\theta = 25^{\circ}$, after the non-linearity of the electro-optic molecules is established by electric field poling. First, we observe at $V_{eo} = 0~$V a shift of the resonant wavelength by 12~nm in the poled sample~($\lambda_{res} = 1540~$nm) compared to the unpoled sample~($\lambda_{res} = 1528~$nm). The experimental Q-factor of the poled sample is $Q=276$. Second, we find that under an applied voltage change from $V_{eo}=100~$V to $V_{eo}=-100~$V, the resonance shifts linearly with applied voltage, as expected (according to $\frac{\Delta \lambda_{res} }{\lambda_{res}}=   -\frac{1}{2}n_{mat}^2r_{33}E\Gamma_c$, with $E=\frac{V_{eo}}{L}$) up to a maximum of $\Delta \lambda_{max}=11$~nm, which suffices to satisfy $\Delta \omega_{eo, 100~V} - \Delta \omega_{eo, -100~V} \sim 2\times \delta \omega_{res} \geq \delta \omega_{res} $~(see inset). We introduce the switching voltage $V_{switch}$ as a figure of merit that quantifies the voltage that is necessary to switch the transmission between its maximal and its minimal value~(which corresponds conceptually to the widely used $V_{\pi}$ in modulators similar to the integrated circuit shown in Fig.~\ref{fig:FigVision}~a). We find a $V_{switch} = 100~V$ to be sufficient to tune the absolute intensity transmitted through the sample at a chosen operation wavelength $\lambda_{OP}$ between its minimum at $T_{min}= 30\%$ and its maximum at $T_{max} = 90\%$ of its maximum value~(shown also in Fig.~\ref{fig:FigResults}~h). This corresponds to a maximal modulation depth $\eta_{max} = \frac{\Delta T}{T_0} = 100\%$, where $\Delta T=T_{max}-T_{min}$ is the total modulation change and $T_0 = \frac{T_{max}+T_{min}}{2}$. In a second example shown in Fig.~\ref{fig:FigResults}~f, we choose to operate on the narrower resonance present when $\theta = 15^{\circ}$ and $\alpha = 0.725$. In this case, we report a maximal tuning of the resonance by $\Delta \lambda_{max}=10$~nm, which corresponds to a $\Delta \omega_{eo, 100~V} - \Delta \omega_{eo, -100~V}  \sim 3.46\times \delta \omega_{res} \geq \delta \omega_{res} $.  Also in this case, at $V_{eo} = 0~$V we observe a shift of the resonant wavelength by 11.6~nm of the poled sample~($\lambda_{res} = 1594~$nm) compared to the unpoled sample~($\lambda_{res} = 1582.4~$nm). The Q-factor of the poled sample is $Q=550$. Importantly, in this case, because of the higher Q-factor, a voltage change of only $V_{eo}=60~$V or $V_{eo}=-60~$V suffices to fully tune through the entire resonance~(see inset).  Consequently, we find that a voltage $V_{switch} = 60~V$ is sufficient switch the absolute intensity transmitted through the sample between its minimum at $60\%$ and its maximum at $100\%$ of its maximum value. In this case,  $\eta_{max} = 50\%$, see Fig.~\ref{fig:FigResults}~h. 

To contrast these two examples, we now analyze in Fig.~\ref{fig:FigResults}~i~-~k the DC tuning behavior when we operate the elliptical resonators on the GMR modes introduced in Fig.~\ref{fig:FigDevice}~d~and~f with geometrical dimensions as provided in the Methods. We find from both experiments and simulations a much broader resonance with an experimental $Q = 37$, where a voltage change of $V_{eo}=100~$V to $V_{eo}=-100~$V tunes the resonant wavelength over a maximal range of $\Delta \lambda_{max} = 20~$nm. This value is approximately twice larger than what we find for the quasi-BIC modes, and can be attributed to a more efficient interaction enabled by the $r_{33}$ electro-optic coefficient due to higher alignment of the nearfield of the nanoresonators with the tuning field (see side-by-side mode profiles and poling/tuning field simulations in Fig.~S2. However, the broad linewidth of the resonance would require a $V_{switch}$ larger than 100~V, thereby demonstrating that GMR can be utilized in scenarios where a large tuning of broad resonances is preferred over a large intensity modulation, as may be the case of modulating broadband emission. Notably, the achieved tuning is approximately twice larger than previous reports that investigated GMR inside hybrid organic-metallic structures from JRD1:PMMA which however did not make use of sub-wavelength resonators~\cite{Benea-ChelmusI.-C.MeretskaM.ElderL.D.TamagnoneM.DaltonR.L.Capasso2021}. 

\begin{figure}[H]
  \centering 
 \includegraphics[width=16cm]{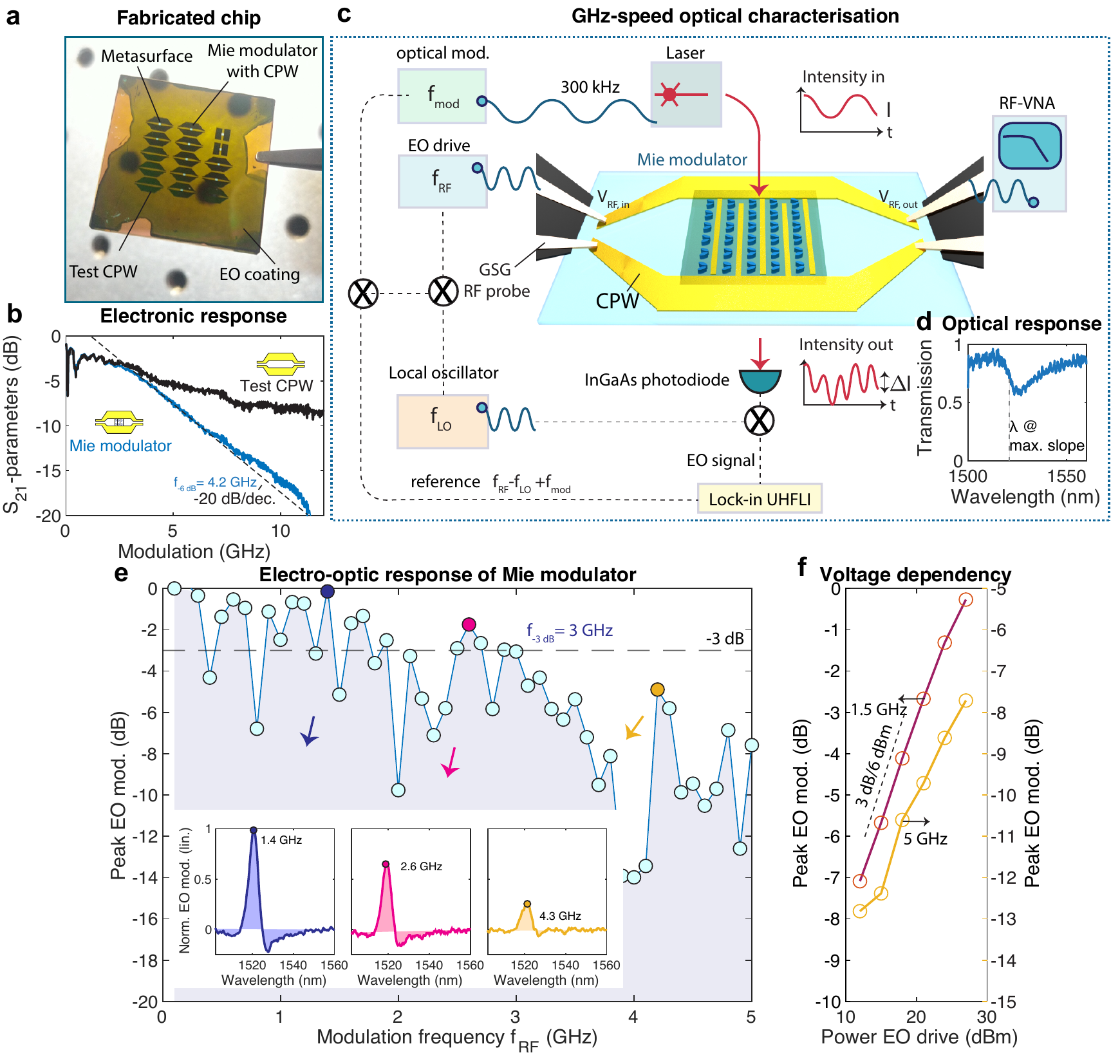}
 \caption{\textcolor{black}{\textbf{GHz-speed properties of the Mie modulators.} \textbf{a,} Picture of a fabricated chip shows Mie modulators that are integrated with GHz coplanar waveguides~(CPW). Also visible are test CPW. \textbf{b,} Electronic scattering parameters $S_{21}$ of Mie modulators are compared to test CPW. The $S_{21}$ are measured using a vector network analyzer~(VNA) connected to the sample by high-frequency cables and high-speed microwave GSG~(ground-source-ground) probes (one ground floating) and exhibits a cut-off of $f_{-6~dB}  = 4.2~$GHz owing to the intrinsic RC bandwidth. RF cable losses are deducted from the $S_{21}$ response. Beyond 4.2~GHz, only the Mie modulators exhibit a decay of -20~dB/decade (much less steep roll-over for the test CPW). \textbf{c,} Optoelectronic experimental setup. The electronic characteristics are measured in a transmission configuration using the VNA, and the wavelength-resolved electro-optical~(EO) modulation is measured using a lock-in amplifier. A double modulation scheme combined with a local oscillator~(LO) is used, where the laser emission is modulated at the source and the Mie modulators~(details in the methods). \textbf{d,} Resonance of sample ($h_{SiO2} = 200~$nm, $\theta = 25^{\circ}$).  \textbf{e,} Peak electro-optic modulation amplitude for frequencies $f_{RF}$ up to 5 GHz. We find a 3-dB electro-optic bandwidth of $f_{EO,-3~dB}  =3$~GHz. Insets: Wavelength-resolved modulation strength for several values of $f_{RF}$, the peak values have been utilized to plot the data in \textbf{e,}. \textbf{f,} Peak electro-optic modulation amplitude for different modulation voltages~(reported as power in dBm), at 1.5 and 5~GHz, the latter well beyond the electro-optic bandwidth.}}
   \label{fig:RFmeasGHz}
\end{figure}

Finally, we analyze the GHz-speed properties of the Mie modulators in Fig.~\ref{fig:RFmeasGHz}. A photograph of several fabricated devices is provided in Fig.~\ref{fig:RFmeasGHz}~a and displays two sets of devices: Mie modulators fitted with interdigitated tuning electrodes that are connected to GHz-speed coplanar waveguides (CPW) and test devices which consist only of the CPW~(no metasurface and no interdigitated electrodes). We first characterize these two structures electrically using a vector network analyzer~(VNA) that outputs the scattering matrix, including the amount of transmitted RF power, characterised by $S_{21, dB}$, using the setup shown in Fig.~\ref{fig:RFmeasGHz}~c. We find a -6~dB cut-off of the Mie modulators at $f_{-6~dB}  = 4.2~$GHz, after which a roll-over of -20~dB/decade is observed, which is agrees well with the RC-time constant of the interdigitated electrode array of the Mie modulators~(see Methods). After the roll-over, the voltage across the modulators drops towards zero. This is in contrast to the test CPW which does not feature such decay. RF cable losses are deducted from the S21 response. Then, we characterize the GHz-speed electro-optic tuning properties of the Mie modulators around their resonance with optical transmission characteristics as shown in Fig.~\ref{fig:RFmeasGHz}~d. We apply a drive field $V_{RF} = V_{eo}\times \sin(2\pi f_{RF}t)$. We use a double modulation scheme in combination with a local oscillator and lock-in detection to characterize the sample up to 5~GHz, above the lock-in bandwidth. Details of the experimental setup are given in the Methods and photographs of the lab setting in Supplementary Fig.~S4.  In Fig.~\ref{fig:RFmeasGHz}~e, we first report the peak electro-optic modulation $\eta_{peak, dB}$ as a function of frequency $f_{RF}$~(we note that here the peak modulation amplitude has been normalized to its value at 100~MHz $\eta_{peak}(f_{RF} = 100~MHz)$ and computed using $\eta_{peak, dB} = 10\log_{10}\frac{\eta_{peak}(f_{RF})}{\eta_{peak}(f_{RF} = 100~MHz)}$). We find that the sample electro-optic bandwidth is $f_{EO,-3~dB}  =3$~GHz,  and that at $f_{RF}=5$~GHz, the modulation amplitude is approximately 7.75~dB lower than the maximum. The discrepancy between the electronic bandwidth and the electro-optic bandwidth can be ascribed to attenuation in the cable from the photodiode to the lock-in amplifier, passing several stages of mixers, which was not accounted for in this experiment. In the inset of Fig.~\ref{fig:RFmeasGHz}~e, we show wavelength-resolved EO modulation for three distinct modulation frequencies~($f_{RF}=~$1.4~GHz, 2.5~GHz and 4.3~GHz at an RF power of 27~dBm at the source). For each wavelength, we normalized the absolute electro-optic modulation to the transmission through the unbiased sample. We find, as expected, that the modulation strength peaks on one side of the asymmetric resonance, more specifically at the wavelength $\lambda$ with the highest slope in the transmission and that it changes sign at the resonance wavelength $\lambda_{res}$. Moreover, we note that a modulation can be measured beyond the 3-dB cut-off, e.g. at $f_{RF}=4.3$~GHz. In Fig.~\ref{fig:RFmeasGHz}~f we investigate the dependence of the modulation amplitude on the drive power at frequencies 1.5~GHz and 5~GHz and observe an approximately linear behavior as expected. 

\section{Discussion and outlook}
Our work using hybrid silicon-organic high-Q metasurfaces and microwave-compatible actuating electrodes is the first step towards a new class of free-space electro-optic modulators that leverage on the unique design flexibility of sub-wavelength resonators covered by an active electro-optic layer to achieve efficient GHz-speed tuning. Also at DC voltages, the use of quasi-BIC modes allows us to achieve a tuning over 31$\%$ of the C-telecom band, corresponding to 11~nm, while GMRs achieve a tuning up to 20~nm, a factor of 2 higher than previous reports that do not use sub-wavelength resonators~\cite{Benea-ChelmusI.-C.MeretskaM.ElderL.D.TamagnoneM.DaltonR.L.Capasso2021}. Furthermore, the demonstrated switching speeds can enable time-dependent and high-speed on-demand control of light, for example for vortex beam generation~\cite{Huang2020,Wang2020} or for time-resolved microscopy and sensing~\cite{Tittl1105}. By reducing the in-plane footprint of the devices from a current approximative area of $330\times330~\mu$m$^2$ to potentially $100\times100~\mu$m$^2$, the intrinsic RC time constant of the device would be reduced by a factor of 10 and would potentially allow operation up to 30~GHz, and represents a key advance for their use as optical links in wireless optical communications. We note here that, recently, bound states in the continuum have been demonstrated even for individual resonators~\cite{Koshelev2020} rather than resonator arrays, thereby suggesting that a further increase in bandwidth beyond tens of GHz might become possible. With the added ability to change the parameters~(angle $\theta$, size) of each single pair of elliptical resonators, the wavefront can be locally affected to achieve an overall functionality of the entire metasurface~\cite{Gigli:21}, for example spatial multiplexing. Importantly, the geometry we propose can be generalized to a much larger variety of Mie resonances, e.g. to achieve phase-only modulation~\cite{Staude2013} or polarisation modulation. Finally, the  achieved relative bandwidth tuning of $0.7\%$ in combination with the high Q-factor and high-speed characteristics may allow to investigate emergent optical phenomena in the area of time-varying~\cite{Shaltout2015} and spatio-temporal~\cite{Wang:20, Shaltouteaat3100} metasurfaces with electro-optic materials~\cite{Wang:20}, and provide an alternative path to magnet-free isolators beyond optomechanical~\cite{Ruesink2016} or piezoelectric~\cite{Tian2021} actuation. Lastly, the device architecture demonstrated here can be used for other nonlinear effects that require highly confined fields in combination with highly performant nonlinear materials, such as second harmonic generation~\cite{Koshelev2020, Anthur2020}, which has so far not been demonstrated in hybrid silicon-organic systems of this kind.

\textbf{Acknowledgments} 
The authors acknowledge insightful discussions with Christopher Bonzon and Michele Tamagnone. I.-C. Benea-Chelmus acknowledges support through the Swiss National Science Foundation for the postdoctoral fellowship P2EZP2.181935 and an independent research grant from the Hans Eggenberger foundation. M. L. Meretska is supported by NWO Rubicon Grant 019.173EN.010, by the Dutch Funding Agency NWO and by the Air Force Office of Scientific Research under award number FA9550-19-1-0352. D.L. Elder and L. R. Dalton acknowledge support from the Air Force Office of Scientific Research (FA9550-19-1-0069). This work was performed in part at the Harvard University Center for Nanoscale Systems (CNS); a member of the National Nanotechnology Coordinated Infrastructure Network (NNCI), which is supported by the National Science Foundation under NSF award no. ECCS-2025158. Additionally, financial support from the Office of Naval Research (ONR) MURI program, under grant no. N00014-20-1-2450 is acknowledged. The electro-optic molecules were synthesized in the Chemistry Department at the University of Washington. 

\textbf{Author contribution} I.-C.B.-C. conceived, designed and implemented the experiments. S.M. assisted with the measurements. I.-C.B.-C. and M.M. fabricated the samples. I.-C.B.-C. did the theoretical derivations and the simulations. D.E. and L.D developed the electro-optic molecules and assisted with the poling of the devices. I.-C.B.-C. and A.S. built the microwave-optical characterization setup. D.K. and A.S. helped with the high-frequency measurements. I.-C.B.-C., M.M., D.K., A.S. and F.C. analysed the data. I.-C.B.-C. wrote the manuscript with feedback from all authors.

\textbf{Competing interests} A provisional patent application with U.S. Serial No.: 63/148,595 has been filed on the subject of this work by the President and Fellows of Harvard College.

\textbf{Corresponding author} Correspondence to Ileana-Cristina Benea-Chelmus (cristinabenea@g.harvard.edu) and Federico Capasso (capasso@seas.harvard.edu).

\textbf{Data availability} The main dataset contained within this paper will be made available in the Zenodo database prior to publication.

\textbf{Code availability} The code used to plot the data within this paper will be made available in the Zenodo database prior to publication.

\bibliography{EOmetasurfaces2}

\section{Methods}

\textbf{Nanofabrication of the hybrid silicon-organic electro-optic modulators.} The structures discussed in this study use in part standard nanofabrication technique used for the silicon-on-insulator platform. These are complemented by a final step in which the organic active layer is applied to the structure and subsequently activated by electric field poling. The fabrication flowchart is shown in Fig,~S1. In short, a multilayer of amorphous silicon~(of thickness 200~nm) on silicon dioxide~(of thickness 200/300~nm) is deposited by chemical vapor deposition~(CVD) onto a quartz substrate. Then, elliptical nanostructures are patterned by electron beam~(e-beam, Elionix 125, 1~$\mu$A current) lithography onto ZEP 520A resist~(spincoated at 3000~rpm) and act as a etch mask in a subsequent two-step fluoride-based reactive ion etching~(RIE, $\mathrm{SF_6, C_4F_8}$), during which the silicon is first etched and then the silicon dioxide is etched using the same resist mask. After etching, the gold electrodes~(15~nm titanium, 35~nm gold) are deposited via e-beam evaporation and subsequent lift-off of ZEP 520A resist in overnight remover PG at 80$^\circ$C. Finally, a mixture of $50\%$~wt JRD1:PMMA~(where PMMA = polymethylmetacrylate) dissolved in 5$\%$wt 1,1,2-trichlorethane is deposited by spin coating at 1000~rpm to the structure to reach a layer thickness of 600-700~nm. After coating, the organic film is dried in an under-vacuum oven at 80~C for 24~hours. After drying, the organic film is rendered electro-optically active by electric field poling, a procedure explained e.g. here~\cite{Xu2020} during which the sample is heated up above the glass temperature of the organic layer~(95~C) and quickly cooled down under an applied poling field on the order of $E_{pol} = 100~V/\mu m$. During this procedure, the electro-optic molecules JRD1 undergo a reorientation from random~(after spin-coating) to being aligned with the poling field lines~(that have a three-dimensional characteristics discussed in greater detail below and in Fig.~S2), owing to their hyperpolarisability. The wavelength-dependent electro-optic properties, the wavelength-dependent and concentration-dependent refractive indices of the JRD1:PMMA mixture are discussed in great detail and provided in the supplementary section of reference~\cite{Benea-ChelmusI.-C.MeretskaM.ElderL.D.TamagnoneM.DaltonR.L.Capasso2021}.

\textbf{Optical and electronic properties of the electro-optic modulators.} The performance of the modulators we demonstrate arises primarily from the fact that the structures we propose support high-Q modes that have a high overlap with the active non-linear material. We compute the quality factor using the formula $Q = \frac{\omega_{res}}{\gamma}$ with $\gamma$ the total loss rate of the resonance which we extract by fitting a Lorentzian lineshape to the transmitted power of the shape $I(\omega) = \frac{A}{(\omega - \omega_{res})^2+(\frac{\gamma}{2})^2}+B$, where A, B, $\omega_{res}$ and $\gamma$ are fitting parameters. In Fig.~S2, we report the simulated distribution of both the optical fields as well as the poling fields (which generate the $\chi^{(2)}$ non-linearity in the organic coating of our hybrid structures). The latter distribution also corresponds closely to the driving DC and RF fields that are applied to the device under test to trigger the electro-optic effect in the modulators. All simulations are done with CST Microwave Studio: the optical simulations are performed using the finite time domain solver, while the DC simulations used the electrostatic solver. The detailed dielectric constant of the organic layer can be found for DC and optical frequencies in reference~\cite{Benea-ChelmusI.-C.MeretskaM.ElderL.D.TamagnoneM.DaltonR.L.Capasso2021}. We first investigate the field distribution at three distinct locations in the plane of the array of resonators: plane A is located 50~nm below the silicon resonators, plane B is located in the center of the silicon resonators and plane C is located 50 nm above the silicon resonators in the JRD1 layer. Furthermore, we provide also the field distribution across the cross-section located at plane D. In all panels, the arrows represent the electric field vector in the considered plane, which the colormap represents the $E_z$-component for the quasi-BIC modes, and the $E_x$-component for the GMR. By analyzing the individual field profiles and their spatial distribution, we notice several notable characteristics of the resonances. Firstly and most importantly, we notice that for both classes of resonances, the optical field is largely concentrated in the nearfield of the silicon resonators and localized in the organic coating. This characteristic ensures a high overlap factor $\Gamma_c$ through a high spatial overlap between the optical mode with the active organic layer, whose refractive index is changed upon an applied driving voltage. Secondly, the field intensity decays rapidly away from the silicon nanopillars and becomes minimal at the location of the interdigitated array of electrodes. Consequently the latter does not affect significantly the linewidth of the resonances. Thirdly, we note that both resonances are excited with x-polarized light. However, the field distribution in the high-intensity nearfield of the resonators is z-polarised for quasi-BIC and x-polarized for GMR. Consequently, an efficient operation of the electro-optic modulator on the $r_{33}$ component is ensured by placing the contact electrodes parallel to the x-axis for the quasi-BIC and parallel to the z-axis for the GMR. The electrodes are then employed to establish the in-device local orientation of the $r_{33}$ electro-optic coefficient. Since the latter aligns with the electric field lines of the poling field, we provide in Fig.~S2 also the electrostatic simulations of the poling field upon an applied voltage for all discussed planes A-D. We find that, as expected, the field amplitude decays far away from the electrodes and as a result, both the local $r_{33}$ coefficient and the total introduced refractive index change $\Delta n$ will be lower at plane C than at plane A. Finally, we underline two important last aspects. First, while the quasi-BIC optical mode is circulating, the GMR mode is mainly linearly polarised, thereby leading to a larger overlap factor with the $r_{33}$ component for the GMR mode and the larger tuning of the resonant wavelength $\Delta \lambda_{res} $ for the GMR. Secondly, because of the high-Q nature of the quasi-BIC modes, an optimal height for the silicon dioxide pedestal was found experimentally to be at 300~nm to ensure at the same time a narrow-band resonance and a large electro-optic tuning at given applied voltage. Increasing the pedestal height would require higher tuning fields to achieve a commensurate effect due to the decay of the poling field as a function of height. 

\textbf{Geometrical properties of silicon nanostructures.} We chose the following dimensions for the quasi-BIC structures: $W = 1.32\alpha$, $L = 1.4~\mu$m,  $D_a = 2\times0.33\alpha$, $D_b = 2\times0.11\alpha$, $d = 0.66\alpha$, $\theta = 15^{\circ},~25^{\circ}$, $\alpha = 0.7$~(Fig.~\ref{fig:FigResults}~a~-~c) and $\alpha = 0.725$~(Fig.~\ref{fig:FigResults}~d~-~f) and number of periods $N_x = 360$ and $N_z = 240$ along the x- and z-axis, respectively. In all cases, the gold electrodes have a width of 200~nm, a length of 330~$\mu$m and a height of 50~nm~(15 nm titanium and 35~nm gold). We chose the following dimensions for the GMR structures: $W = 1.4~\mu$m, $L = 1.32\alpha$,  $D_a = 2\times0.33\alpha$, $D_b = 2\times0.11\alpha$, $d = 0.66\alpha$, $\theta = 15^{\circ}$, number of periods $N_x = 240$ and $N_z = 360$ and $\alpha = 0.75$. Also, $h_{SiO2} = 200~$nm. For the high-frequency measurements shown in Fig.~\ref{fig:RFmeasGHz}, we used a quasi-BIC structure with $\alpha = 0.675$. 

\textbf{High-frequency optoelectronic characterisation of Mie modulators.} First, we characterize the electronic $S_{21, dB} = 20 \log_{10}\frac{V_{RF, out}}{V_{RF,in}}$ parameters using a vector network analyzer~(VNA, Agilent E8364B) by contacting two GSG probes from GGB~(Picoprobe 40A series, DC to 40~GHz) to the CPW of the modulators. Second, a double modulation scheme~($f_{mod}$, $f_{EO}$) in combination with a local oscillator~(LO, $f_{LO}$) is used to characterize the high-frequency electro-optic tuning properties of the Mie modulators. The laser is internally modulated at $f_{mod} = 300$~kHz, 50~$\%$ duty cycle and full intensity modulation using an external source~(pulser Agilent B114A). The electro-optic modulators are modulated at the speeds reported in Fig.~\ref{fig:RFmeasGHz} using a second RF source that outputs a sinusoidal signal~(Hittite Microwave Corporation HMC-T2100B, 10~MHz to 20~GHz). The modulated laser intensity is detected by a photodiode~(Newport 1544-A, bandwidth 12~GHz). A third RF source~(Wiltron Anritsu 68347B, 10~MHz to 20~GHz) is used as a local oscillator at frequency $f_{LO} = f_{RF}+41$~MHz to mix down the photodiode signal to an intermediate frequency of $f_{IF} = 41~$MHz, irrespective of the modulating frequencies $f_{RF}$~(mixer ZMF-2-S+, bandwidth 1-1000~MHz and mixer ZX05-C42-S+, bandwidth 1000-4200 MHz, both from Mini-circuits). The downmixed the photodiode signal is recorded by a high-frequency lock-in amplifier~(UHFLI from Zurich instruments, with maximal demodulation frequency 600~MHz). A second set of mixers~(mixer ZMF-2-S+ from Mini-circuits, bandwidth 1-1000~MHz and mixer ZLW-1-1+ from Mini-circuits, bandwidth 0.1-500~MHz) is used to mix the intermediate frequency with the modulation frequency to form the reference signal at $f_{ref} = f_{IF} + f_{mod}$ that is used to demodulate the downmixed photodiode signal and to report the modulated intensity. This double modulation scheme combined with the local oscillator is necessary to unambiguously detect the electro-optic modulation of the sample in frequency ranges that are larger than the cut-off of the lock-in amplifier. The light incident onto the Mie modulators is collimated prior to the sample (diameter 6~mm) and then focussed onto the sample using a lens with focal length 100~mm. The sample is placed in the focus of the beam and its position is adjusted using an xyz stage. A linear polarizer filters any polarization components that are not parallel to the x-axis. 

\textbf{RC time constant of Mie modulators.} The Mie modulators investigated in this work have dimensions as discussed above. Their switching speed is mainly determined by the capacitance of the interdigitated array of electrodes, loaded with the organic material and with silicon pillars, and by the 50~$\Omega$ resistance of the source. We consider a simplified model~\cite{OLTHUIS1995252} for the computation of the capacitance per period, per unit length, using the formula below:

\begin{equation}
C_{per}/L = \frac{\epsilon_0(\epsilon_{SiO_2}+\epsilon_{JRD1:PMMA})}{4}\frac{K(\sqrt{1-k^2})}{K(k)}
\end{equation}

where $\epsilon_{SiO_2} = 3.75$, $\epsilon_{JRD1:PMMA} = 5$, $k = \cos(\frac{\pi w_{electrodes}}{2w_{gap}})$ and $w_{electrodes}=0.2~\mu$m is the width of the interdigitated electrodes and $w_{gap} = 1.2~\mu$m, the gap between two electrodes. $K(k)= \int_{0}^{1}\frac{dt}{[(1-t^2)(1-k^2t^2)]^{0.5}}$ are the elliptical integrals of first kind. At a total length of $L = 300~\mu$m and a total number of periods of $N_z = 240$, the total capacitance is equal to $C_{tot} = N_z\times C_{per}/L \times L = 0.27$~pF. With this, the RC cut-off frequency of the device is estimated analytically by assuming $R = R_{source}+R_{device}$~($R_{source} = 50~\Omega$ and $R_{device} = 24~\Omega$ the serial resistance of the device, measured on samples where all interdigitated electrodes were shorted) to $f_{-3~dB, calc} = 2.6~$GHz and $f_{-6~dB, calc} = 4.5~$GHz. We note however that there is a variance in the device resistance owing to the quality of the gold wires which might affect in turn the cut-off frequency. This formula does not consider the elliptical silicon resonators, which increase the capacitance compared to this estimated value and hence lower the RC cut-off frequency. 

\end{document}